\documentclass[english]{article}
\usepackage[utf8]{inputenc}
\usepackage{rotfloat}
\usepackage{amsmath}
\usepackage{longtable}
\usepackage{tabularx}
\usepackage{float}
\usepackage{color, colortbl}
\usepackage[letterpaper, margin=1in]{geometry}
\usepackage{multirow}
\usepackage{placeins}
\usepackage{babel}
\usepackage{color}
\usepackage{advdate}

\pdfoutput=1
\usepackage{hyperref}
\hypersetup{
  pdfinfo={
    Title={Your Title Here},
    Author={Your Name Here},
    Subject={If you want to put something here, do so},
    Keywords={Add some keywords if you feel so inclined}
  }}
\usepackage{pdfpages}

\makeatletter
\date{\AdvanceDate[-57]\today}

\makeatother

\begin{document}

\newgeometry{left=2.54cm, right=2.54cm, top=4.5cm}

\title{Response of Energy Levels and B(E2)s to Variations of\\
Single Particle Energies\\
\vspace{1.4em}}

\author{Castaly Fan$^{1}$, Praveen C. Srivastava$^{2}$, and Larry Zamick$^{1}$ \\
\vspace{1.4em}
\\
$^{1}$Department of Physics and Astronomy, Rutgers University, \\
 Piscataway, New Jersey 08854, USA\\
$^{2}$Department of Physics, Indian Institute of Technology \\
 Roorkee, Roorkee 247667, India}
 
\maketitle

\begin{abstract}

    We perform shell model calculations for $^{48}$Cr using GXPF1 interaction by varying single particle energies.  First we keep the splitting of the pair A ($p_{3/2} - f_{7/2}$) constant and likewise  pair B ($p_{1/2} - f_{5/2}$). We then shift the pair B energy by an amount  $\Delta$  relative to the first pair. We then study the dependence of the energies of the even J states in $^{48}$Cr. We further study the effects on B(E2)'s $J$ to $J-2$.  We then do a similar analysis with pair C ($f_{5/2} -f_{7/2}$) constant  and we shift pair D ($p_{1/2}-p_{3/2}$) by an amount  $\Delta$ .  In the latter case the spin orbit splittings are held constant. Again we study the effects on energies and B(E2)s. We also discuss the possibility of scaling.

\end{abstract}

\section{Introduction}

    In this work we examine the effects of varying single particle energies in a shell model calculation using the GXFP1A interaction \cite{1}. We will focus on the yrast spectrum of $^{48}$Cr $J=0, 2, 4,..., 16$. We perform calculations with the GXFP1A interaction \cite{1}. There have been many publications using these interactions in the PF shell with the shell model program ANTOINE \cite{3}, notably the works of E. Caurier et al. \cite{4}\cite{5} and more recently of V. Kumar et al. \cite{6}. Other works related to $^{48}$Cr include those of K. Hara et al. \cite{7}, F.Brandolini et al. \cite{8}, E.Caurier et al. \cite{9}, Z.C.Gao et al. \cite{10} and R.A. Herrera et al. \cite{11}.

\clearpage

\restoregeometry

\section{Arrangement of Tables and Figures}

    The original single particle energies are given in the second column of Table \ref{t1}. In the third column we show the corresponding values for the FPD6 interaction \cite{2}. The orbits are 0$f_{7/2} , 1p_{3/2}, 0f_{5/2}$ and $1p_{1/2}$. We will then alter the single particle energies as indicated in Tables \ref{t2}, \ref{t3}, \ref{t4}, and \ref{t5}. We first perform calculations with what we call CASE-1.
    
    \begin{table}[H]
    \centering
    \caption{Single particle energies (MeV) of GXPF1A and FPD6.}
    \vspace{0.2cm}
    \begin{tabular}{|c|c|c|}
    \hline 
    Orbit  & GXPF1A  & FPD6\tabularnewline
    \hline 
    \hline 
    $0f_{7/2}$  & 0 (-8.6240)  & 0 (-8.3876)\tabularnewline
    \hline 
    $1p_{3/2}$  & 2.9447  & 1.8942\tabularnewline
    \hline 
    $0f_{5/2}$  & 7.2411  & 6.4910\tabularnewline
    \hline 
    $1p_{1/2}$  & 4.4870  & 3.9093\tabularnewline
    \hline 
    \end{tabular}
    \label{t1}
    \end{table}

    In Table \ref{t2} we keep the splitting of the pair A ($p_{3/2} - f_{7/2}$) constant and likewise  pair B ($p_{1/2} - f_{5/2}$). We then shift the pair B energy by a positive  amount $\Delta$ relative to the first pair. We then study the dependence of the energies of the even $J$ states in $^{48}$Cr. We further study the effects on B(E2)'s $J$ to $J-2$.  In Table \ref{t3} we extend the calculation to negative $\Delta$. 

    Next we perform calculations  which what we call CASE-2. In Tables \ref{t4} and \ref{t5} do a similar analysis as was done in CASE-1, but now with pair C ($f_{5/2} -f_{7/2}$) constant and we pair D ($p_{1/2}-p_{3/2}$) shifted  by an amount $\Delta$.  In the latter case the spin orbit splittings are held constant. Again we study the effects on energies and B(E2)s.

\begin{sidewaystable}[H]
    \caption{Energy spectra of $^{48}$Cr using GXPF1A interaction. Here we have keep the single-particle energies of $0f_{7/2}$
    and $1p_{3/2}$ as the original one, and changed the single-particle
    energies of $0f_{5/2}$ and $1p_{1/2}$ moved up by original plus $\Delta$.}
    \label{t2}
    \resizebox{\textwidth}{6cm}{%
    \begin{tabular}{|c|c|c|c|c|c|c|c|c|c|c|c|c|c|c|c|c|}
    \hline 
    Energy  & GXPF1A  & $\Delta$ = 1  & 2  & 3  & 4  & 5  & 6  & 7  & 8  & 9  & 10  & 20  & 40  & 60  & 80  & 100 \tabularnewline
    \hline 
    $0^{+}$  & 0.000  & 0.000  & 0.000  & 0.000  & 0.000  & 0.000  & 0.000  & 0.000  & 0.000  & 0.000  & 0.000  & 0.000  & 0.000  & 0.000  & 0.000  & 0.000 \tabularnewline
    \hline 
    $2^{+}$  & 0.788  & 0.767  & 0.750  & 0.737  & 0.727  & 0.718  & 0.711  & 0.705  & 0.700  & 0.696  & 0.692  & 0.669  & 0.654  & 0.646  & 0.644  & 0.642 \tabularnewline
    \hline 
    $4^{+}$  & 1.717  & 1.687  & 1.664  & 1.646  & 1.632  & 1.620  & 1.610  & 1.601  & 1.594  & 1.588  & 1.582  & 1.547  & 1.519  & 1.508  & 1.501  & 1.498 \tabularnewline
    \hline 
    $6^{+}$  & 3.229  & 3.152  & 3.090  & 3.039  & 2.998  & 2.963  & 2.934  & 2.908  & 2.886  & 2.867  & 2.849  & 2.747  & 2.671  & 2.639  & 2.622  & 2.610 \tabularnewline
    \hline 
    $8^{+}$  & 4.753  & 4.649  & 4.553  & 4.478  & 4.415  & 4.361  & 4.314  & 4.274  & 4.239  & 4.208  & 4.179  & 4.001  & 3.881  & 3.827  & 3.796  & 3.778 \tabularnewline
    \hline 
    $10^{+}$  & 6.429  & 6.238  & 6.080  & 5.952  & 5.846  & 5.758  & 5.684  & 5.621  & 5.565  & 5.517  & 5.474  & 5.219  & 5.033  & 4.957  & 4.915  & 4.889 \tabularnewline
    \hline 
    $12^{+}$  & 7.722  & 7.479  & 7.296  & 7.153  & 7.037  & 6.941  & 6.860  & 6.791  & 6.731  & 6.679  & 6.632  & 6.356  & 6.153  & 6.070  & 6.024  & 5.996 \tabularnewline
    \hline 
    $14^{+}$  & 9.701  & 9.432  & 9.227  & 9.063  & 8.929  & 8.818  & 8.724  & 8.693  & 8.572  & 8.511  & 8.456  & 8.129  & 7.887  & 7.788  & 7.733  & 7.699 \tabularnewline
    \hline 
    $16^{+}$  & 12.805  & 12.411  & 12.115  & 11.845  & 11.699  & 11.546  & 11.417  & 11.308  & 11.213  & 11.130  & 11.057  & 10.623  & 10.305  & 10.156  & 10.105  & 10.061 \tabularnewline
    \hline 
    \hline 
    B(E2: J $\rightarrow$ J-2 ) ($e^{2}fm^{4}$)  &  &  &  &  &  &  &  &  &  &  &  &  &  &  &  & \tabularnewline
    \hline 
    $2^{+}$  & 249  & 237  & 228  & 222  & 217  & 212  & 209  & 206  & 204  & 202  & 200  & 191  & 184  & 180  & 179  & 178 \tabularnewline
    \hline 
    $4^{+}$  & 336  & 321  & 310  & 302  & 296  & 291  & 287  & 283  & 280  & 278  & 276  & 262  & 252  & 248  & 245  & 244 \tabularnewline
    \hline 
    $6^{+}$  & 336  & 317  & 304  & 296  & 290  & 285  & 280  & 277  & 274  & 271  & 270  & 256  & 247  & 244  & 241  & 240 \tabularnewline
    \hline 
    $8^{+}$  & 306  & 288  & 276  & 268  & 262  & 257  & 253  & 250  & 248  & 245  & 244  & 232  & 223  & 222  & 218  & 217 \tabularnewline
    \hline 
    $10^{+}$  & 212  & 195  & 192  & 186  & 186  & 185  & 185  & 184  & 184  & 184  & 183  & 181  & 180  & 179  & 179  & 178 \tabularnewline
    \hline 
    $12^{+}$  & 162  & 163  & 162  & 162  & 162  & 161  & 161  & 160  & 160  & 160  & 159  & 157  & 156  & 155  & 154  & 154 \tabularnewline
    \hline 
    $14^{+}$  & 126  & 125  & 124  & 123  & 123  & 123  & 123  & 122  & 122  & 121  & 121  & 120  & 118  & 118  & 118  & 117 \tabularnewline
    \hline 
    $16^{+}$  & 62  & 65  & 66  & 67  & 68  & 68  & 68  & 68  & 68  & 68  & 68  & 68  & 68  & 67  & 67  & 67 \tabularnewline
    \hline 
    \hline 
    Q$(eb)$  &  &  &  &  &  &  &  &  &  &  &  &  &  &  &  & \tabularnewline
    \hline 
    $2^{+}$  & -0.30  & -0.29  & -0.29  & -0.28  & -0.28  & -0.27  & -0.27  & -0.27  & -0.27  & -0.27  & -0.26  & -0.26  & -0.25  & -0.24  & -0.25  & -0.25 \tabularnewline
    \hline 
    $4^{+}$  & -0.40  & -0.39  & -0.38  & -0.38  & -0.36  & -0.36  & -0.36  & -0.35  & -0.35  & -0.34  & -0.34  & -0.33  & -0.31  & -0.31  & -0.30  & -0.29 \tabularnewline
    \hline 
    $6^{+}$  & -0.40  & -0.38  & -0.37  & -0.36  & -0.35  & -0.34  & -0.33  & -0.33  & -0.32  & -0.32  & -0.32  & -0.29  & -0.28  & -0.28  & -0.27  & -0.27 \tabularnewline
    \hline 
    $8^{+}$  & -0.41  & -0.38  & -0.36  & -0.34  & -0.33  & -0.32  & -0.31  & -0.30  & -0.29  & -0.28  & -0.28  & -0.26  & -0.22  & -0.24  & -0.22  & -0.22 \tabularnewline
    \hline 
    $10^{+}$  & -0.21  & -0.17  & -0.16  & -0.15  & -0.13  & -0.13  & -0.12  & -0.12  & -0.12  & -0.11  & -0.11  & -0.10  & -0.09  & -0.09  & -0.09  & -0.09 \tabularnewline
    \hline 
    $12^{+}$  & -0.03  & -0.02  & -0.02  & -0.02  & -0.02  & -0.02  & -0.02  & -0.02  & -0.02  & -0.02  & -0.02  & -0.02  & -0.02  & -0.02  & -0.02  & -0.02 \tabularnewline
    \hline 
    $14^{+}$  & -0.05  & -0.04  & -0.03  & -0.03  & -0.03  & -0.03  & -0.03  & -0.03  & -0.02  & -0.02  & -0.02  & -0.02  & -0.01  & -0.01  & -0.01  & -0.01 \tabularnewline
    \hline 
    $16^{+}$  & -0.09  & -0.08  & -0.07  & -0.07  & -0.06  & -0.06  & -0.06  & -0.06  & -0.05  & -0.05  & -0.05  & -0.04  & -0.04  & -0.04  & -0.04  & -0.04 \tabularnewline
    \hline 
    \end{tabular} }
\end{sidewaystable}

\begin{sidewaystable}[H]
    \caption{Energy spectra of $^{48}$Cr using GXPF1A interaction. Here
    we have keep the single-particle energies of $0f_{7/2}$
    and $1p_{3/2}$ as the original one, and changed the single-particle
    energies of $0f_{5/2}$ and $1p_{1/2}$ moved up by original minus $\Delta$.}
    \label{t3}
    
    \resizebox{\textwidth}{6cm}{%
    \begin{tabular}{|c|c|c|c|c|c|c|c|c|c|c|c|c|c|c|c|}
    \hline 
    Energy  & $\Delta$ = -1  & -2  & -3  & -4  & -5  & -6  & -7  & -8  & -9  & -10  & -20  & -40  & -60  & -80  & -100 \tabularnewline
    \hline 
    $0^{+}$  & 0.000  & 0.000  & 0.000  & 0.000  & 0.000  & 0.000  & 0.000  & 0.000  & 0.000  & 0.000  & 0.000  & 0.000  & 0.000  & 0.000  & 0.000 \tabularnewline
    \hline 
    $2^{+}$  & 0.816  & 0.845  & 0.865  & 0.872  & 0.878  & 0.882  & 0.876  & 0.857  & 0.842  & 0.910  & 0.490  & 0.385  & 0.359  & 0.349  & 0.340 \tabularnewline
    \hline 
    $4^{+}$  & 1.757  & 1.808  & 1.861  & 1.915  & 1.968  & 1.995  & 1.973  & 1.890  & 1.760  & 1.710  & 1.424  & 1.134  & 1.061  & 1.025  & 1.007 \tabularnewline
    \hline 
    $6^{+}$  & 3.316  & 3.376  & 3.352  & 3.359  & 3.433  & 3.496  & 3.494  & 3.410  & 3.234  & 1.864  & 2.650  & 2.153  & 2.036  & 1.9165  & 1.931 \tabularnewline
    \hline 
    $8^{+}$  & 4.877  & 4.970  & 4.987  & 5.056  & 5.177  & 5.261  & 5.260  & 5.147  & 4.626 & 4.058  & 2.966  & 2.433  & 2.297  & 2.237  & 2.202 \tabularnewline
    \hline 
    $10^{+}$  & 6.621  & 6.663  & 6.633  & 6.760  & 6.965  & 7.102  & 7.114  & 6.998  & 6.439  & 5.770  & 7.675  & 6.865  & 6.651  & 6.552  & 6.495 \tabularnewline
    \hline 
    $12^{+}$  & 8.060  & 8.539  & 9.099  & 9.397  & 9.469  & 9.449  & 9.373  & 9.131  & 8.195  & 7.487  & 15.588  & 35.059  & 54.929  & 74.871  & 94.838 \tabularnewline
    \hline 
    $14^{+}$  & 10.060  & 10.547  & 11.133  & 11.585  & 11.648  & 11.589  & 11.512  & 11.397  & 10.959  & 10.434  & 26.249  & 65.601  & 105.436  & 145.363  & 185.319 \tabularnewline
    \hline 
    $16^{+}$  & 13.361  & 14.034  & 14.169  & 13.720  & 13.424  & 13.263  & 13.134  & 12.965  & 12.778  & 12.854  & 28.411  & 67.637  & 107.439  & 147.349  & 187.297 \tabularnewline
    \hline 
    \hline 
    B(E2: J $\rightarrow$ J-2 ) ($e^{2}fm^{4}$)  &  &  &  &  &  &  &  &  &  &  &  &  &  &  & \tabularnewline
    \hline 
    $2^{+}$  & 268  & 295  & 326  & 348  & 359  & 362  & 360  & 356  & 343  & 306  & 141  & 128  & 124  & 123  & 122 \tabularnewline
    \hline 
    $4^{+}$  & 362  & 400  & 449  & 484  & 499  & 501  & 492  & 457  & 388  & 242  & 182  & 165  & 161  & 158  & 156 \tabularnewline
    \hline 
    $6^{+}$  & 369  & 421  & 448  & 466  & 491  & 511  & 516  & 501  & 429  & 172  & 151  & 137  & 114  & 135  & 130 \tabularnewline
    \hline 
    $8^{+}$  & 344  & 418  & 487  & 523  & 552  & 568  & 571  & 540  & 205 & 199  & 8  & 8  & 21  & 6  & 7 \tabularnewline
    \hline 
    $10^{+}$  & 249  & 317 & 382  & 428  & 476  & 511  & 517  & 400  & 223  & 191  & 0.12  & 0.06  & 0.06  & 0.05  & 0.05 \tabularnewline
    \hline 
    $12^{+}$  & 155  & 141  & 236  & 360  & 412  & 440  & 446  & 15  & 198  & 173  & 0.008  & 0.009  & 0.009  & 0.009  & 0.009 \tabularnewline
    \hline 
    $14^{+}$  & 131  & 147  & 212  & 324  & 390  & 403  & 389  & 0.45  & 6  & 2  & 0.028  & 0.02  & 0.02  & 0.02  & 0.02 \tabularnewline
    \hline 
    $16^{+}$  & 56  & 24  & 98  & 114  & 209  & 245  & 255  & 246  & 75  & 46  & 0.23  & 0.73  & 0.96  & 0.94  & 1.093 \tabularnewline
    \hline 
    \hline 
    Q$(eb)$  &  &  &  &  &  &  &  &  &  &  &  &  &  &  & \tabularnewline
    \hline 
    $2^{+}$  & -0.32  & -0.32  & -0.31  & -0.29  & -0.27  & -0.26  & -0.26  & -0.26  & -0.26  & -0.18  & +0.24  & +0.22  & +0.22  & +0.23  & +0.22 \tabularnewline
    \hline 
    $4^{+}$  & -0.42  & -0.43  & -0.43  & -0.41  & -0.41  & -0.41  & -0.42  & -0.45  & -0.49  & -0.51  & +0.30  & +0.29  & +0.28  & +0.28  & +0.28 \tabularnewline
    \hline 
    $6^{+}$  & -0.42  & -0.42  & -0.37  & -0.34  & -0.34  & -0.35  & -0.37  & -0.41  & -0.54  & -0.61  & +0.29  & +0.26  & +0.13  & +0.27  & +0.24 \tabularnewline
    \hline 
    $8^{+}$  & -0.43  & -0.43  & -0.40  & -0.39  & -0.38  & -0.38  & -0.38  & -0.46  & -0.70  & -0.68  & -0.40  & -0.38  & -0.38  & -0.38  & -0.38 \tabularnewline
    \hline 
    $10^{+}$  & -0.27  & -0.37  & -0.41  & -0.42  & -0.42  & -0.42  & -0.41  & -0.42  & -0.74  & -0.72  & -0.01  & -0.005  & -0.003  & -0.002  & -0.0019 \tabularnewline
    \hline 
    $12^{+}$  & -0.05  & -0.10  & -0.30  & -0.42  & -0.42  & -0.41  & -0.40  & -0.75  & -0.74  & -0.72  & -0.48  & -0.46  & -0.45  & -0.45  & -0.45 \tabularnewline
    \hline 
    $14^{+}$  & -0.08  & -0.16  & -0.33  & -0.44  & -0.43  & -0.42  & -0.41  & -0.40  & -0.49  & -0.48  & -0.42  & -0.39  & -0.39  & -0.39  & -0.39 \tabularnewline
    \hline 
    $16^{+}$  & -0.01  & -0.32  & -0.36  & -0.43  & -0.44  & -0.44  & -0.43  & -0.43  & -0.44  & -0.49  & -0.51  & -0.48  & -0.48  & -0.47  & -0.47 \tabularnewline
    \hline 
    \end{tabular}}
\end{sidewaystable}

\begin{sidewaystable}[H]
    \caption{Energy spectra of $^{48}$Cr using GXPF1A interaction. Here
    we have keep the single-particle energies of $0f_{7/2}$
    and $0f_{5/2}$ as the original one, and changed the single-particle
    energies of $1p_{3/2}$ and $1p_{1/2}$ moved up by original plus $\Delta$.}
    \label{t4}
    
    \resizebox{\textwidth}{6cm}{%
    \begin{tabular}{|c|c|c|c|c|c|c|c|c|c|c|c|c|c|c|c|}
    \hline 
    Energy  & $\Delta$ = 1  & 2  & 3  & 4  & 5  & 6  & 7  & 8  & 9  & 10  & 20  & 40  & 60  & 80  & 100 \tabularnewline
    \hline 
    $0^{+}$  & 0.000  & 0.000  & 0.000  & 0.000  & 0.000  & 0.000  & 0.000  & 0.000  & 0.000  & 0.000  & 0.000  & 0.000  & 0.000  & 0.000  & 0.000 \tabularnewline
    \hline 
    $2^{+}$  & 0.918  & 1.008  & 1.068  & 1.107  & 1.133  & 1.152  & 1.165  & 1.175  & 1.182  & 1.187  & 1.205  & 1.207  & 1.206  & 1.206  & 1.205 \tabularnewline
    \hline 
    $4^{+}$  & 1.803  & 1.864  & 1.907  & 1.937  & 1.960  & 1.976  & 1.989  & 1.999  & 2.007  & 2.014  & 2.041  & 2.051  & 2.053  & 2.054  & 2.054 \tabularnewline
    \hline 
    $6^{+}$  & 3.293  & 3.312  & 3.311  & 3.303  & 3.293  & 3.283  & 3.274  & 3.265  & 3.257  & 3.250  & 3.205  & 3.171  & 3.157  & 3.150  & 3.145 \tabularnewline
    \hline 
    $8^{+}$  & 4.729  & 4.699  & 4.671  & 4.647  & 4.625  & 4.607  & 4.591  & 4.578  & 4.565  & 4.555  & 4.491  & 4.444  & 4.425  & 4.416  & 4.409 \tabularnewline
    \hline 
    $10^{+}$  & 6.198  & 6.040  & 5.940  & 5.864  & 5.807  & 5.762  & 5.725  & 5.696  & 5.670  & 5.649  & 6.352  & 5.455  & 5.425  & 5.409  & 5.399 \tabularnewline
    \hline 
    $12^{+}$  & 7.396  & 7.108  & 6.948  & 6.833  & 6.747  & 6.680  & 6.626  & 6.582  & 6.544  & 6.513  & 6.346  & 6.241  & 6.201  & 6.179  & 6.167 \tabularnewline
    \hline 
    $14^{+}$  & 9.244  & 8.954  & 8.757  & 8.614  & 8.507  & 8.423  & 8.355  & 8.300  & 8.253  & 8.213  & 8.002  & 7.869  & 7.819  & 7.792  & 7.776 \tabularnewline
    \hline 
    $16^{+}$  & 12.280  & 11.939  & 11.702  & 11.529  & 11.396  & 11.292  & 11.207  & 11.137  & 11.078  & 11.027  & 10.754  & 10.578  & 10.511  & 10.475  & 10.453 \tabularnewline
    \hline 
    \hline 
    B(E2: J $\rightarrow$ J-2 ) ($e^{2}fm^{4}$)  &  &  &  &  &  &  &  &  &  &  &  &  &  &  & \tabularnewline
    \hline 
    $2^{+}$  & 213  & 189  & 173  & 161  & 153  & 147  & 142  & 138  & 135  & 133  & 121  & 115  & 112  & 111  & 111 \tabularnewline
    \hline 
    $4^{+}$  & 276  & 234  & 205  & 186  & 171  & 161  & 153  & 148  & 143  & 139  & 122  & 114  & 110  & 109  & 109 \tabularnewline
    \hline 
    $6^{+}$  & 268  & 224 & 196  & 178  & 166  & 156  & 150  & 145  & 140  & 137  & 122  & 115  & 111  & 111  & 111 \tabularnewline
    \hline 
    $8^{+}$  & 255  & 223  & 203  & 190  & 180  & 173  & 168  & 164  & 160  & 157  & 143  & 135  & 132  & 131  & 131 \tabularnewline
    \hline 
    $10^{+}$  & 184  & 170  & 161  & 155 & 150  & 147  & 143  & 140  & 138  & 135  & 127  & 121  & 119  & 118  & 117 \tabularnewline
    \hline 
    $12^{+}$  & 153  & 145  & 139  & 134  & 131  & 127  & 125  & 123  & 121  & 119  & 112  & 107  & 105  & 103  & 103 \tabularnewline
    \hline 
    $14^{+}$  & 120  & 114  & 110  & 107  & 104  & 102  & 100  & 99  & 98  & 97  & 90  & 87  & 85  & 85  & 84 \tabularnewline
    \hline 
    $16^{+}$  & 60  & 59  & 57  & 56  & 55  & 55  & 54  & 54  & 53  & 53  & 50  & 48  & 48  & 47  & 47 \tabularnewline
    \hline 
    \hline 
    Q$(eb)$  &  &  &  &  &  &  &  &  &  &  &  &  &  &  & \tabularnewline
    \hline 
    $2^{+}$  & -0.26  & -0.22  & -0.18  & -0.16  & -0.13  & -0.12  & -0.10  & -0.09  & -0.08  & -0.08  & -0.04  & -0.01  & -0.0091  & -0.0068  & -0.0047 \tabularnewline
    \hline 
    $4^{+}$  & -0.36  & -0.34  & -0.32  & -0.30  & -0.28  & -0.27  & -0.27  & -0.26  & -0.25  & -0.25  & -0.22  & -0.19  & -0.19  & -0.18  & -0.18 \tabularnewline
    \hline 
    $6^{+}$  & -0.32  & -0.26  & -0.22  & -0.19  & -0.17  & -0.15  & -0.14  & -0.13  & -0.12  & -0.12  & -0.08  & -0.06  & -0.05  & -0.05  & -0.05 \tabularnewline
    \hline 
    $8^{+}$  & -0.33  & -0.27  & -0.23  & -0.20  & -0.18  & -0.17  & -0.15  & -0.15  & -0.13  & -0.13  & -0.09  & -0.08  & -0.07  & -0.070  & -0.06 \tabularnewline
    \hline 
    $10^{+}$  & -0.15  & -0.12  & -0.10  & -0.09  & -0.08  & -0.08  & -0.07  & -0.07  & -0.06  & -0.06  & +0.01  & -0.04  & -0.04  & -0.041  & -0.04 \tabularnewline
    \hline 
    $12^{+}$  & -0.02  & -0.02  & -0.02  & -0.02  & -0.01  & -0.01  & -0.01  & -0.01  & -0.01  & -0.01  & -0.01  & -0.009  & -0.009  & -0.008  & -0.008 \tabularnewline
    \hline 
    $14^{+}$  & -0.04  & -0.04  & -0.03  & -0.03  & -0.03  & -0.03  & -0.03  & -0.03  & -0.03  & -0.02  & -0.02  & -0.02  & -0.02  & -0.02  & -0.02 \tabularnewline
    \hline 
    $16^{+}$  & -0.06  & -0.05  & -0.04  & -0.03  & -0.03  & -0.03  & -0.03  & -0.03  & -0.02  & -0.02  & -0.02  & -0.02  & -0.01  & -0.01  & -0.01 \tabularnewline
    \hline 
    \end{tabular}} 
\end{sidewaystable}

\begin{sidewaystable}[H]
    \caption{Energy spectra of $^{48}$Cr using GXPF1A interaction. Here
    we have keep the single-particle energies of $0f_{7/2}$
    and $0f_{5/2}$ as the original one, and changed the single-particle
    energies of $0p_{3/2}$ and $1p_{1/2}$ moved up by original minus $\Delta$.}
    \label{t5}
    
    \resizebox{\textwidth}{6cm}{%
    \begin{tabular}{|c|c|c|c|c|c|c|c|c|c|c|c|c|c|c|c|}
    \hline 
    Energy  & $\Delta$ = -1  & -2  & -3  & -4  & -5  & -6  & -7  & -8  & -9  & -10  & -20  & -40  & -60  & -80  & -100 \tabularnewline
    \hline 
    $0^{+}$  & 0.000  & 0.000  & 0.000  & 0.000  & 0.000  & 0.000  & 0.000  & 0.000  & 0.000  & 0.000  & 0.000  & 0.000  & 0.000  & 0.000  & 0.000 \tabularnewline
    \hline 
    $2^{+}$  & 0.638  & 0.532  & 0.540  & 0.795  & 1.232  & 1.294  & 1.317  & 1.333  & 1.344  & 1.352  & 1.367  & 1.361  & 1.358  & 1.355  & 1.353 \tabularnewline
    \hline 
    $4^{+}$  & 1.612  & 1.539  & 1.584  & 1.881  & 2.738  & 3.838  & 4.334  & 4.446  & 4.501  & 4.529  & 4.552  & 4.520  & 4.505  & 4.497  & 4.491 \tabularnewline
    \hline 
    $6^{+}$  & 3.103  & 2.991  & 3.061  & 3.577  & 5.074  & 7.084  & 8.712  & 9.843  & 10.860  & 11.862  & 21.830  & 41.799  & 61.787  & 81.781  & 101.778 \tabularnewline
    \hline 
    $8^{+}$  & 4.753  & 4.745  & 4.863  & 5.389  & 6.785  & 8.699  & 10.701  & 12.712  & 14.683  & 15.949  & 25.960  & 45.866  & 65.829  & 85.809  & 105.797 \tabularnewline
    \hline 
    $10^{+}$  & 6.728  & 6.848  & 6.935  & 7.594  & 9.491  & 11.692  & 13.547  & 15.467  & 17.414  & 19.376  & 39.219  & 79.143  & 119.119  & 159.105  & 199.098 \tabularnewline
    \hline 
    $12^{+}$  & 8.341  & 8.942  & 8.984  & 9.706  & 11.763  & 14.550  & 17.441  & 20.075  & 22.152  & 24.139  & 43.927  & 83.780  & 123.727  & 163.699  & 203.682 \tabularnewline
    \hline 
    $14^{+}$  & 10.459  & 11.392  & 11.312  & 12.011  & 14.324  & 17.686  & 21.251  & 24.414  & 27.380  & 30.332  & 60.070  & 119.925  & 179.874  & 239.848  & 299.832 \tabularnewline
    \hline 
    $16^{+}$  & 13.655  & 14.610  & 14.605  & 15.296  & 17.590  & 20.955  & 24.656  & 28.478  & 32.358  & 36.270 & 75.918  & 155.754  & 235.698  & 315.670  & 395.653 \tabularnewline
    \hline 
    \hline 
    B(E2: J $\rightarrow$ J-2 ) ($e^{2}fm^{4}$)  &  &  &  &  &  &  &  &  &  &  &  &  &  &  & \tabularnewline
    \hline 
    $2^{+}$  & 302  & 359  & 388  & 363  & 257  & 190  & 170  & 158  & 151 & 145  & 123  & 112  & 109  & 107  & 106 \tabularnewline
    \hline 
    $4^{+}$  & 417  & 496  & 523  & 413  & 65  & 13  & 222  & 188  & 167  & 154  & 116  & 104  & 100  & 99  & 98 \tabularnewline
    \hline 
    $6^{+}$  & 425  & 500  & 508  & 425  & 273  & 189  & 22  & 18  & 44 & 51  & 54  & 54  & 54  & 55  & 54 \tabularnewline
    \hline 
    $8^{+}$  & 385  & 454  & 453  & 368  & 234  & 118  & 8  & 0.04  & 0.60  & 3  & 2 & 2  & 2  & 2  & 1 \tabularnewline
    \hline 
    $10^{+}$  & 293  & 391  & 392  & 319  & 195  & 3  & 2 & 2  & 4  & 3  & 0.45  & 0.32  & 0.29  & 0.28 & 0.27 \tabularnewline
    \hline 
    $12^{+}$  & 171  & 316  & 341  & 325  & 282  & 24  & 21  & 18  & 13  & 11  & 10  & 10  & 10  & 9  & 9 \tabularnewline
    \hline 
    $14^{+}$  & 139  & 232  & 267  & 261  & 210  & 153  & 136  & 56  & 22  & 15  & 8  & 7  & 7  & 7  & 7 \tabularnewline
    \hline 
    $16^{+}$  & 66  & 132  & 160  & 159  & 156  & 154  & 125  & 43  & 23  & 17  & 9  & 8  & 7  & 7  & 7 \tabularnewline
    \hline 
    \hline 
    Q$(eb)$  &  &  &  &  &  &  &  &  &  &  &  &  &  &  & \tabularnewline
    \hline 
    $2^{+}$  & -0.35  & -0.39  & -0.40  & -0.35  & +0.13  & +0.25  & +0.26  & +0.25  & +0.25  & +0.25  & +0.23  & +0.22  & +0.22  & +0.21  & +0.21 \tabularnewline
    \hline 
    $4^{+}$  & -0.45  & -0.50  & -0.53  & -0.55  & -0.53  & -0.50  & +0.29  & +0.32  & +0.32  & +0.32  & +0.30  & +0.28  & +0.28  & +0.28  & +0.27 \tabularnewline
    \hline 
    $6^{+}$  & -0.49  & -0.56  & -0.61  & -0.63  & -0.61  & -0.49  & -0.14  & +0.01  & +0.07  & +0.09  & +0.12  & +0.13  & +0.13  & +0.13  & +0.13 \tabularnewline
    \hline 
    $8^{+}$  & -0.51  & -0.60  & -0.67  & -0.70  & -0.69  & -0.67  & -0.65  & -0.63  & -0.54  & -0.13  & -0.09  & -0.09  & -0.09  & -0.09  & -0.08 \tabularnewline
    \hline 
    $10^{+}$  & -0.38  & -0.61  & -0.69  & -0.69  & -0.66  & -0.35  & -0.34  & -0.33  & -0.32  & -0.32  & -0.29  & -0.27  & -0.27  & -0.26  & -0.26 \tabularnewline
    \hline 
    $12^{+}$  & -0.08  & -0.59  & -0.69  & -0.69  & -0.67  & -0.63  & -0.58  & -0.37  & -0.28  & -0.26  & -0.24  & -0.24  & -0.24  & -0.23  & -0.24 \tabularnewline
    \hline 
    $14^{+}$  & -0.08  & -0.57  & -0.70  & -0.68  & -0.66  & -0.64  & -0.58  & -0.48  & -0.46  & -0.45  & -0.41  & -0.40  & -0.39  & -0.39  & -0.39 \tabularnewline
    \hline 
    $16^{+}$  & -0.14  & -0.59  & -0.71  & -0.70  & -0.68  & -0.66  & -0.65  & -0.64  & -0.63  & -0.62 & -0.59  & -0.58  & -0.57  & -0.57  & -0.57 \tabularnewline
    \hline 
    \end{tabular}} 
\end{sidewaystable}

\begin{table}[H]
    \centering
    \caption{Ratio E(J)$_{\Delta}$/E(J) with GXPF1A for $\Delta$=1,10, and 20.}
    \label{t6} 
    \vspace{0.3cm}
    \begin{tabular}{|c|c|c|c|}
    \hline 
    $J/\Delta$  & 1  & 10  & 20\tabularnewline
    \hline 
    \hline 
    2  & .973  & .878  & .849\tabularnewline
    \hline 
    4  & .983  & .921  & .901\tabularnewline
    \hline 
    6  & .976  & .882  & .851\tabularnewline
    \hline 
    8  & .978  & .879  & .842\tabularnewline
    \hline 
    10  & .977  & .851  & .812\tabularnewline
    \hline 
    12  & .968  & .859  & .823\tabularnewline
    \hline 
    14  & .972  & .872  & .838\tabularnewline
    \hline 
    16  & .969  & .863  & .829\tabularnewline
    \hline 
    Q(2$^{+})$  & .967  & .867  & .867\tabularnewline
    \hline 
    $\sqrt{B(E2)}$  & .975  & 0.896  & .876\tabularnewline
    \hline 
    \end{tabular}.
\end{table}

\begin{table}[H]
    \centering
    \caption{Comparison of original energies with normalized ones for $\Delta=$20 using GXPF1A. Renormalization factor = 1.2.}
    \label{t7} 
    \vspace{0.3cm}
    \begin{tabular}{|c|c|c|}
    \hline 
    $J$  & Original Spectrum  & $\Delta$=20 Renormalized\tabularnewline
    \hline 
    \hline 
    0  & 0.000  & 0.000\tabularnewline
    \hline 
    2  & 0.788  & 0.802\tabularnewline
    \hline 
    4  & 1.717  & 1.854\tabularnewline
    \hline 
    6  & 3.279  & 3.294?\tabularnewline
    \hline 
    8  & 4.752  & 4.801\tabularnewline
    \hline 
    10  & 6.420  & 6.268?\tabularnewline
    \hline 
    12  & 7.722  & 7.627\tabularnewline
    \hline 
    14  & 9.701  & 9.755\tabularnewline
    \hline 
    16  & 12.805  & 12.748\tabularnewline
    \hline
    \end{tabular}
\end{table}

\clearpage

    We now discuss the arrangements of the figures. There are 4 sets, each with 4 subsets:

    \subsection{SET 1}
    \subsection*{CASE - 1}
    Pair B ($f_{5/2}-p_{1/2}$) shifted by an amount $\Delta$ relative to Pair A ($p_{3/2}-f_{7/2}$).
    \begin{figure}[H]
    \centering
        \begin{minipage}{0.5\textwidth}
        \centering
        \includegraphics[width=1\textwidth]{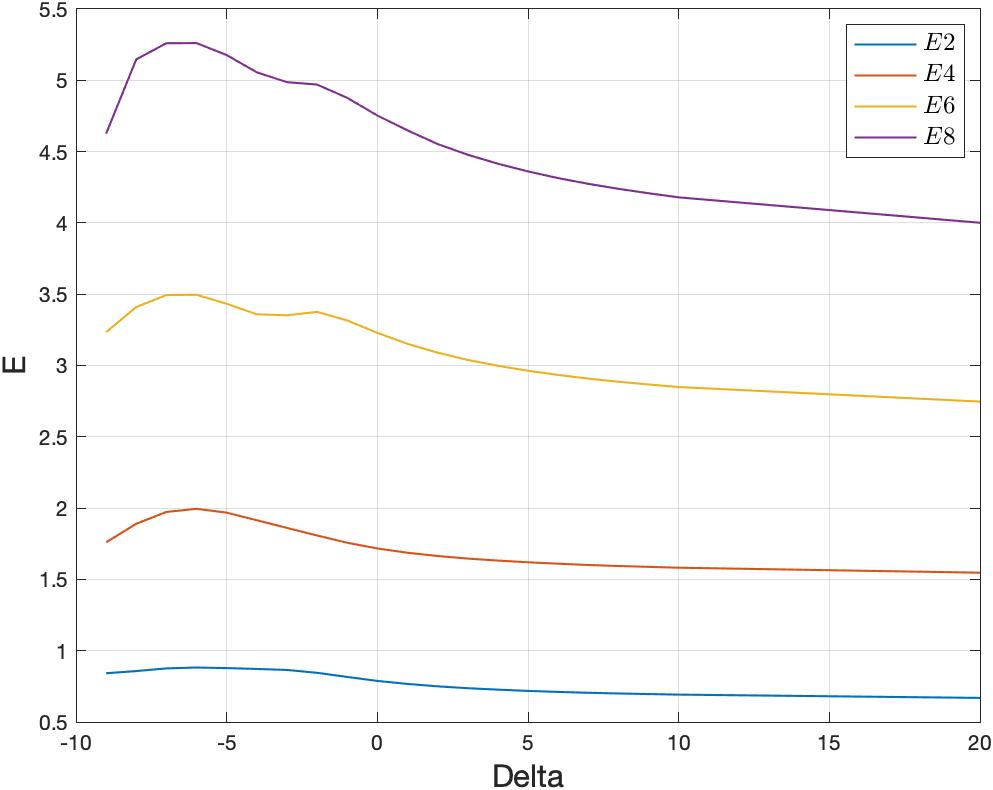}
        \caption{E2, E4, E6, E8.}
        \label{f11}
        \end{minipage}%
        \begin{minipage}{0.5\textwidth}
        \centering
        \includegraphics[width=1\textwidth]{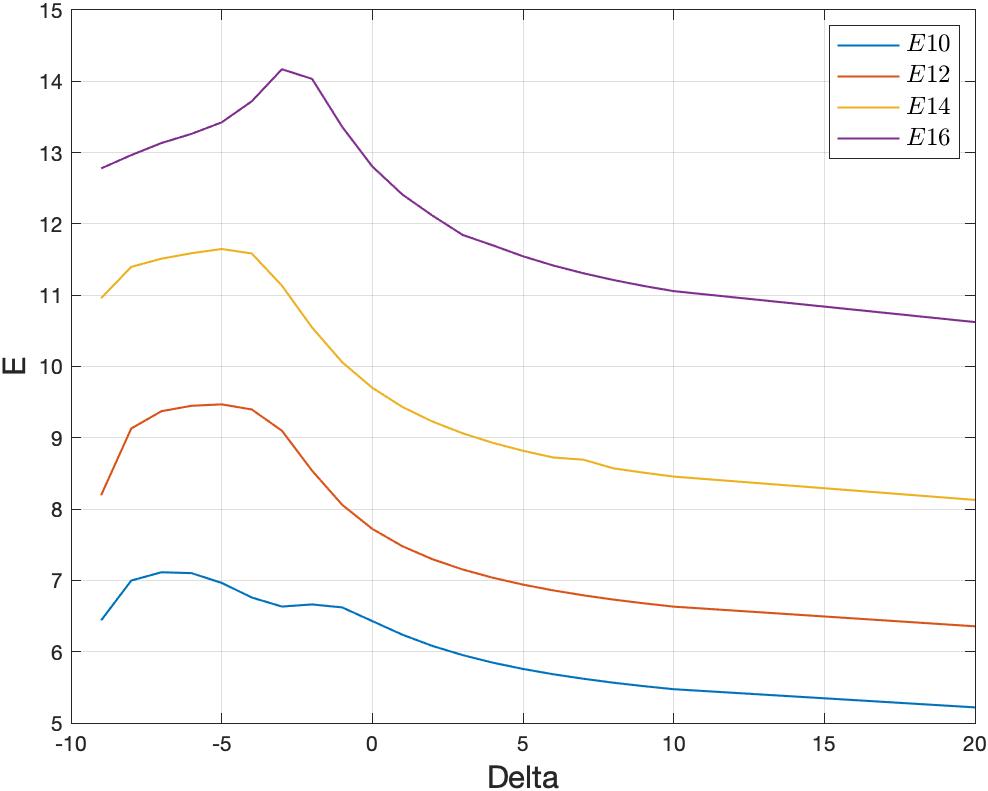}
        \caption{E10, E12, E14, E16.}
        \label{f12}
        \end{minipage}
    \end{figure}
    
    \begin{figure}[H]
    \centering
        \begin{minipage}{0.5\textwidth}
        \centering
        \includegraphics[width=1\textwidth]{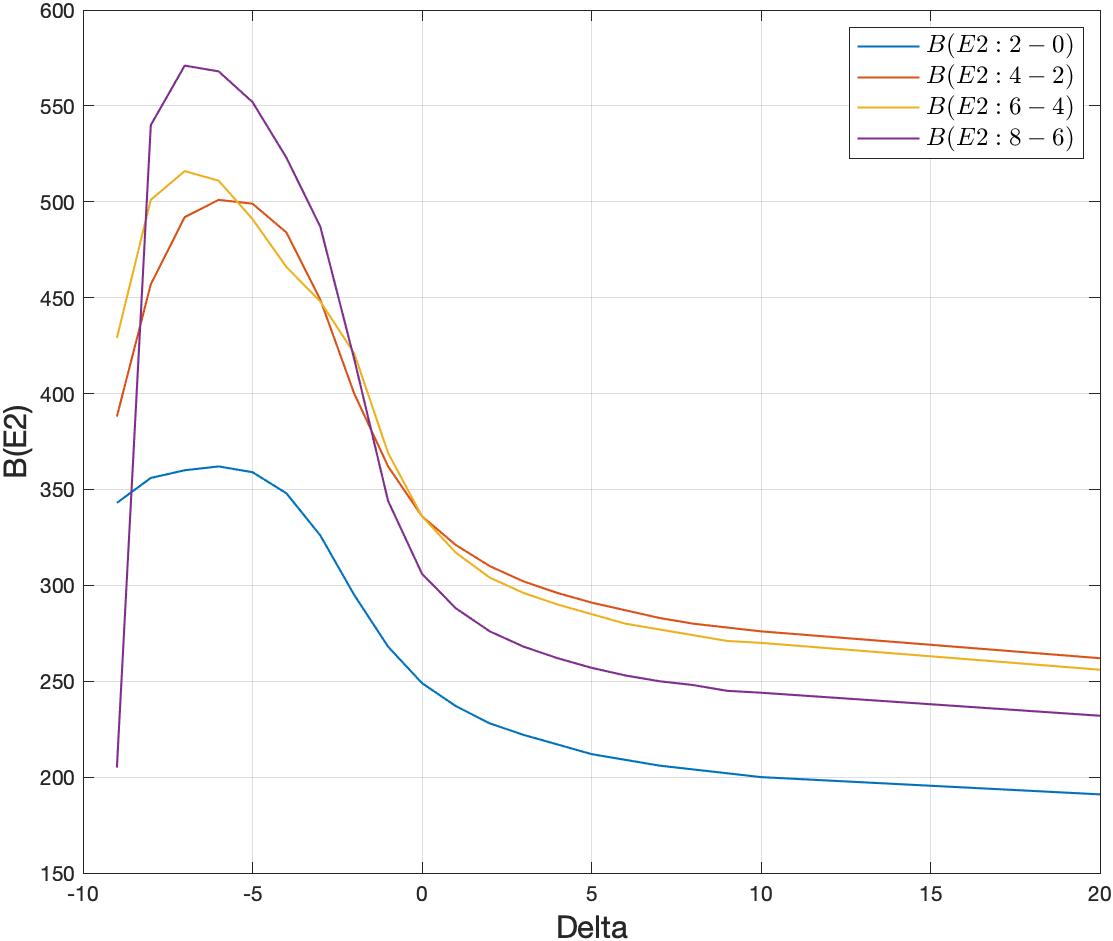}
        \caption{B(E2) 2, 4, 6, 8.}
        \label{f13}
        \end{minipage}%
        \begin{minipage}{0.5\textwidth}
        \centering
        \includegraphics[width=1\textwidth]{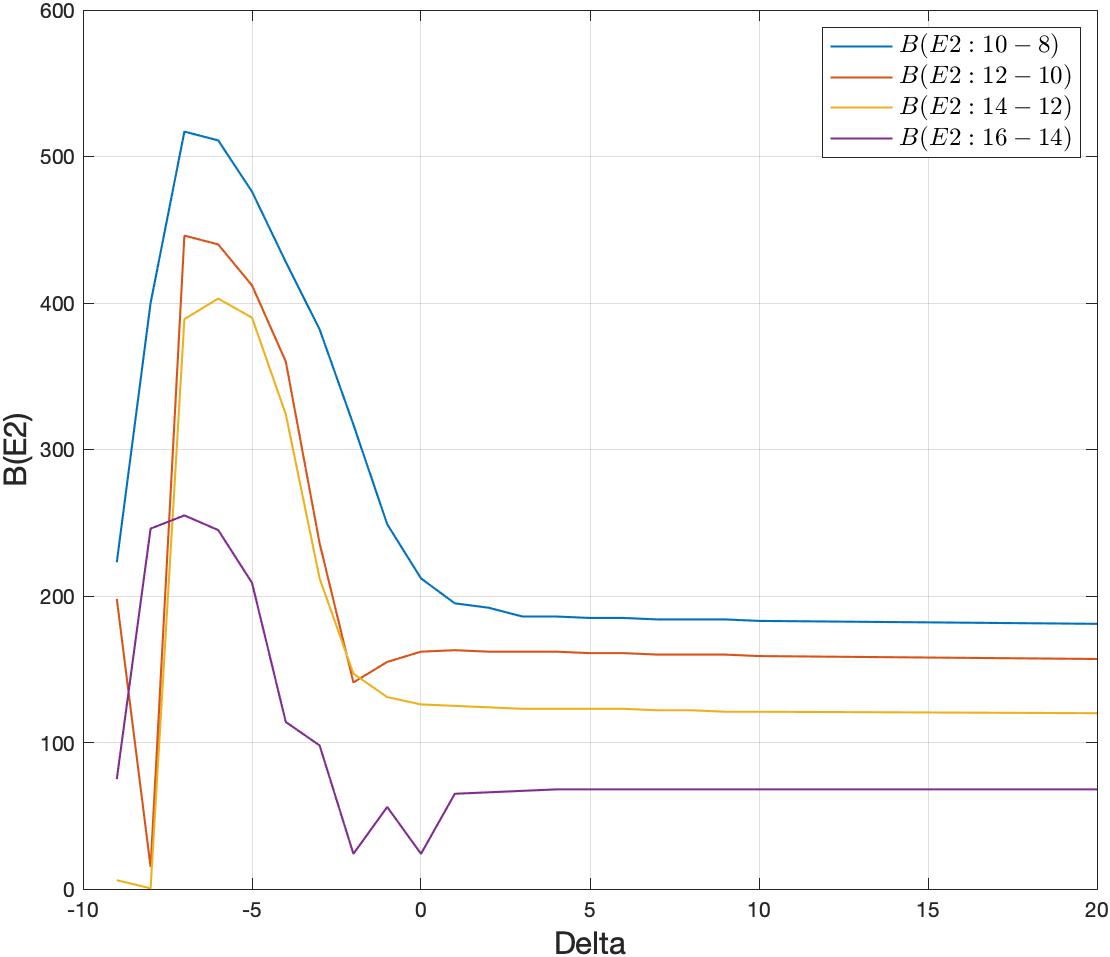}
        \caption{B(E2) 10, 12, 14, 16.}
        \label{f14}
        \end{minipage}
    \end{figure}

    \clearpage

    \subsection{SET 2}
    \subsection*{CASE - 2}
    Pair D ($p_{1/2}-p_{3/2}$) shifted by an amount $\Delta$ relative to pair D ($f_{5/2}-f_{7/2}$).
    \begin{figure}[H]
    \centering
        \begin{minipage}{0.5\textwidth}
        \centering
        \includegraphics[width=1\textwidth]{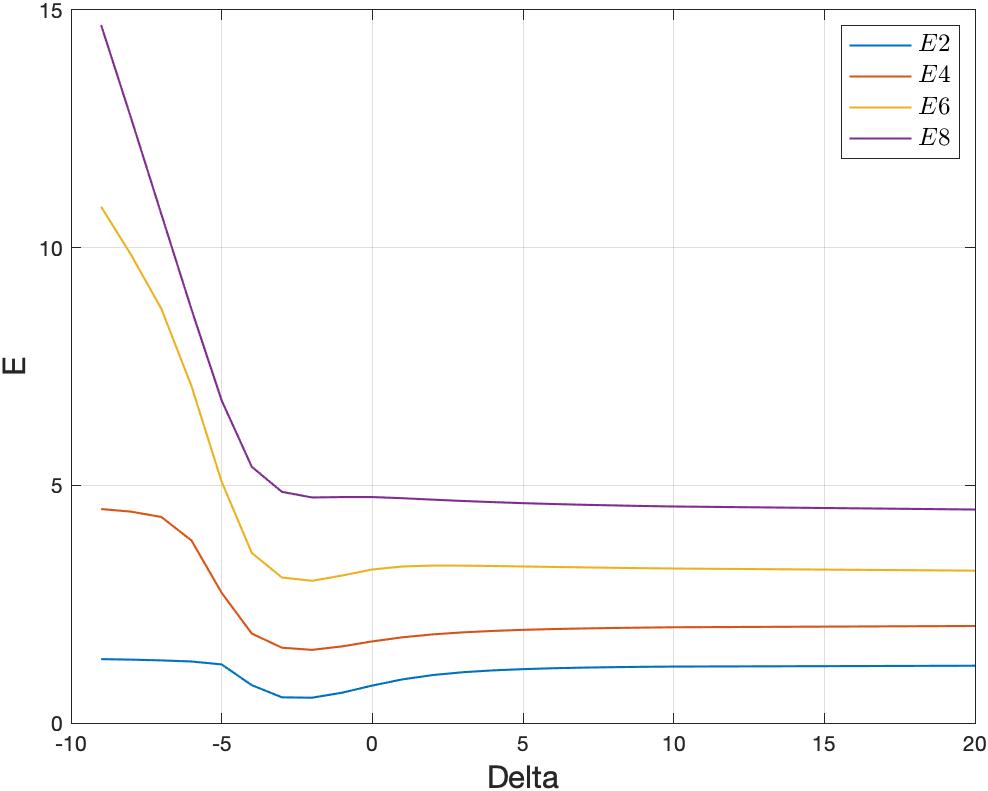}
        \caption{E2, E4, E6, E8.}
        \label{f21}
        \end{minipage}%
        \begin{minipage}{0.5\textwidth}
        \centering
        \includegraphics[width=1\textwidth]{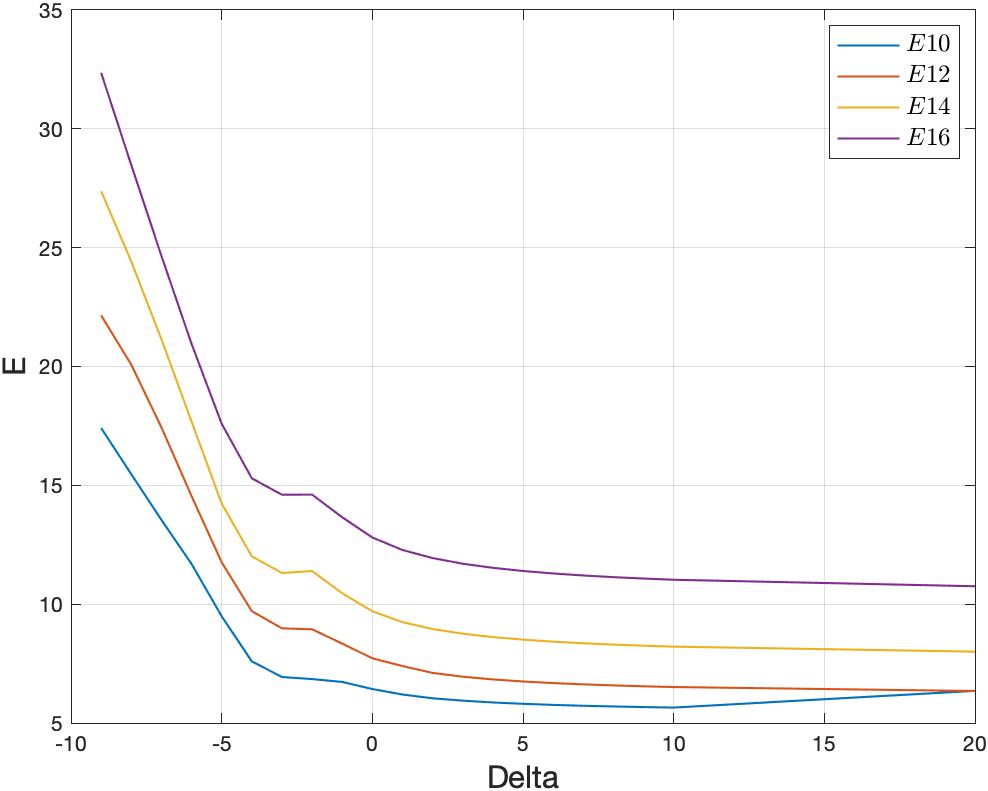}
        \caption{E10, E12, E14, E16.}
        \label{f22}
        \end{minipage}
    \end{figure}
    
    \begin{figure}[H]
    \centering
        \begin{minipage}{0.5\textwidth}
        \centering
        \includegraphics[width=1\textwidth]{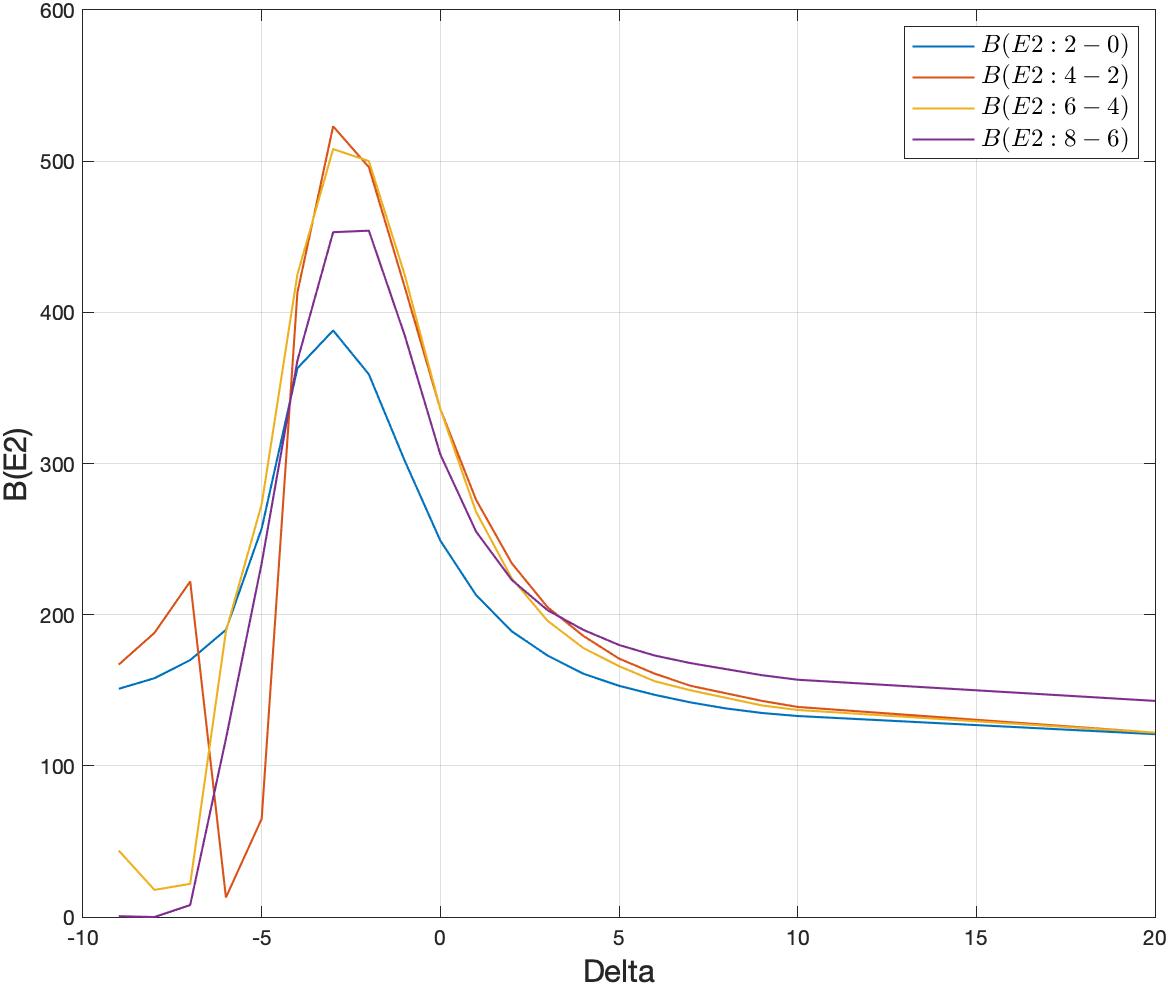}
        \caption{B(E2) 2, 4, 6, 8.}
        \label{f23}
        \end{minipage}%
        \begin{minipage}{0.5\textwidth}
        \centering
        \includegraphics[width=1\textwidth]{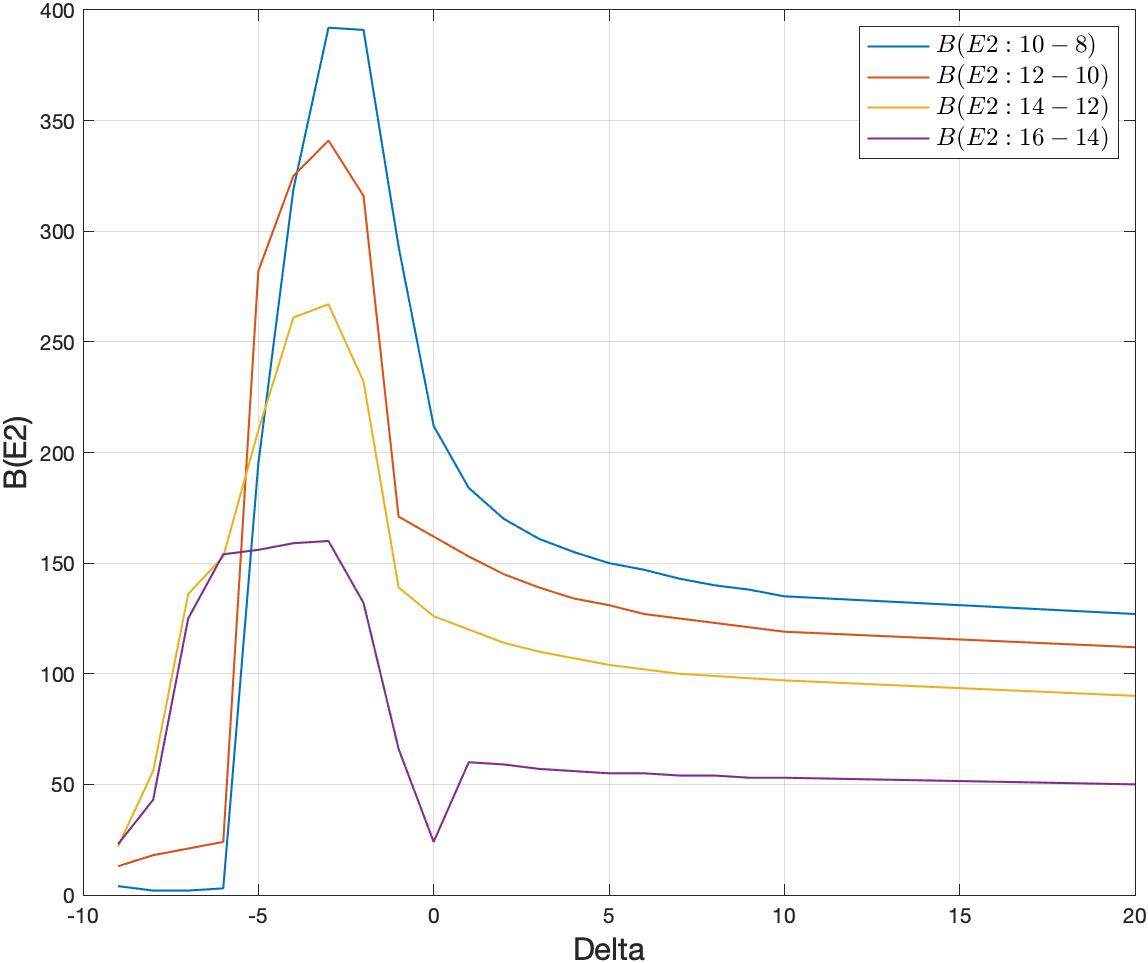}
        \caption{B(E2) 10, 12, 14, 16.}
        \label{f24}
        \end{minipage}
    \end{figure}
    
    \clearpage

    \subsection{SET 3}
    \subsection*{CASE - 1}
    Global view: Pair B ($f_{5/2}-p_{1/2}$) shifted by an amount $\Delta$ relative to Pair A ($p_{3/2}-f_{7/2}$).
    \begin{figure}[H]
    \centering
        \begin{minipage}{0.5\textwidth}
        \centering
        \includegraphics[width=1\textwidth]{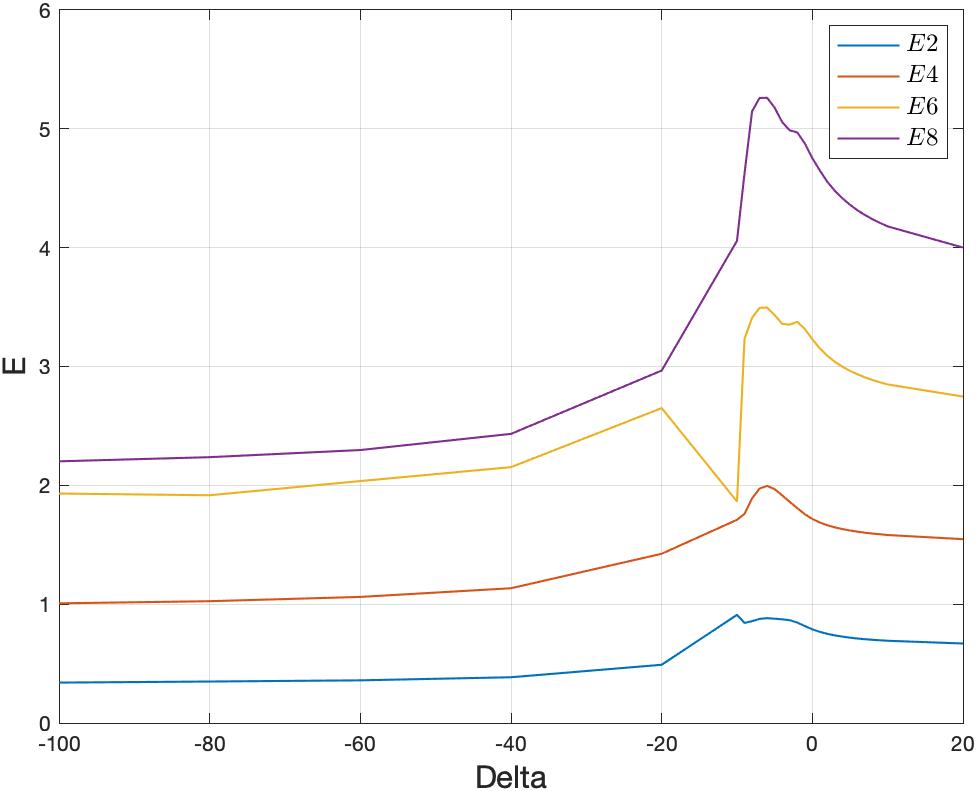}
        \caption{E2, E4, E6, E8.}
        \label{f31}
        \end{minipage}%
        \begin{minipage}{0.5\textwidth}
        \centering
        \includegraphics[width=1\textwidth]{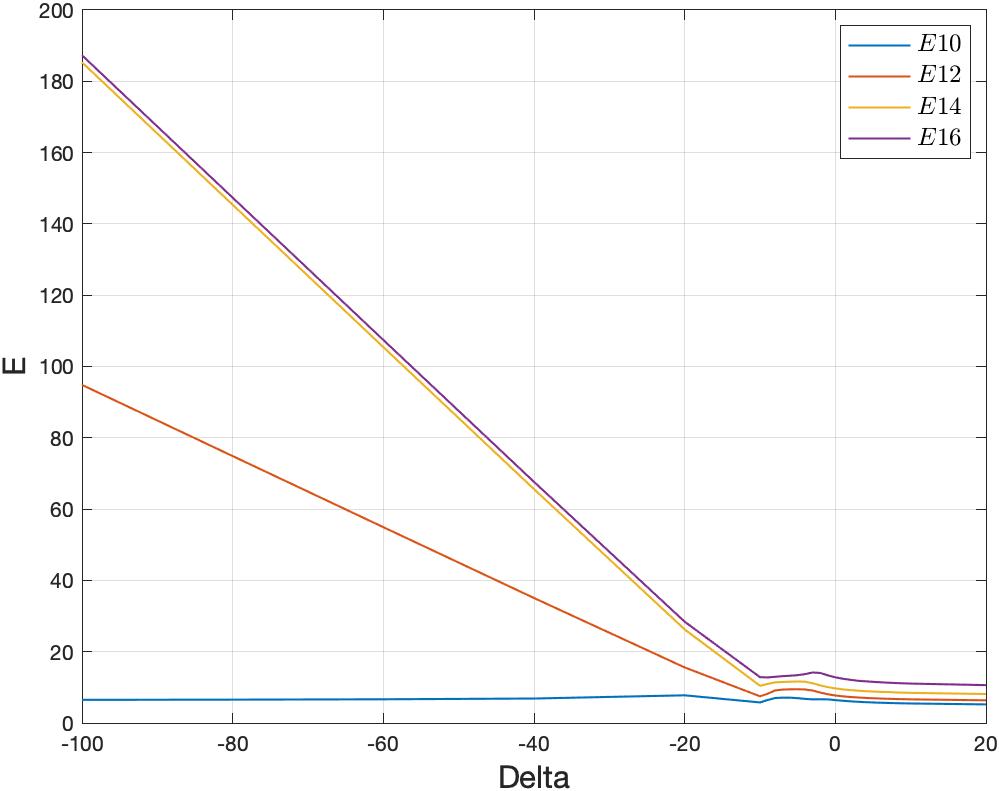}
        \caption{E10, E12, E14, E16.}
        \label{f32}
        \end{minipage}
    \end{figure}
    
    \begin{figure}[H]
    \centering
        \begin{minipage}{0.5\textwidth}
        \centering
        \includegraphics[width=1\textwidth]{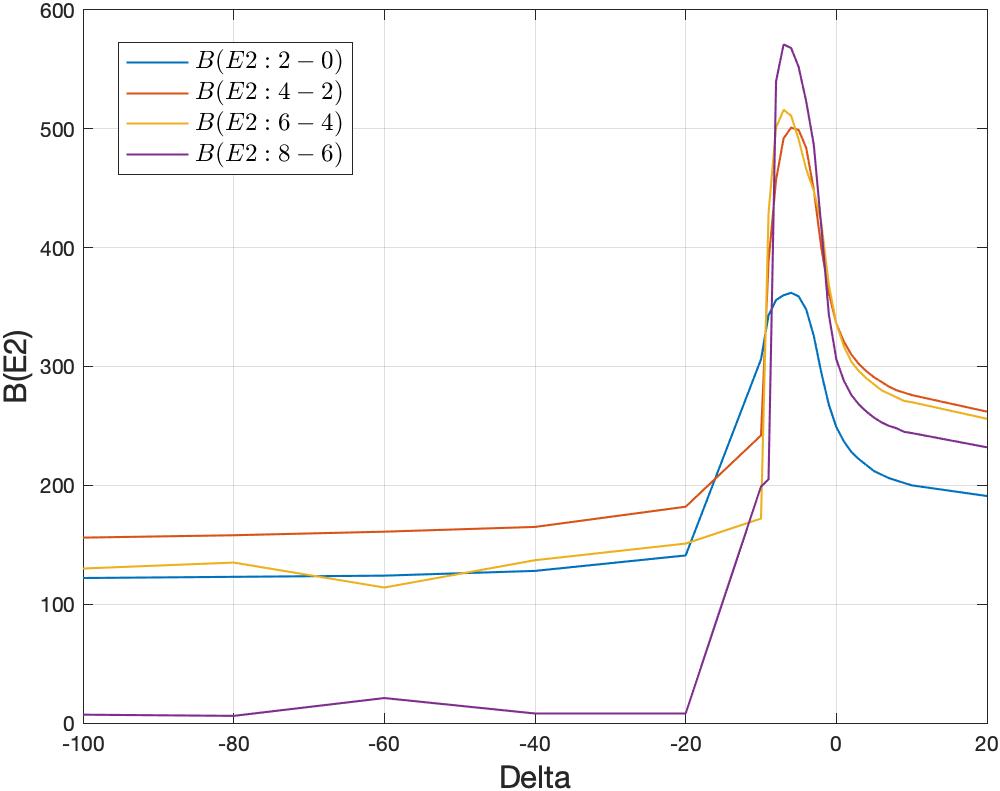}
        \caption{B(E2) 2, 4, 6, 8.}
        \label{f33}
        \end{minipage}%
        \begin{minipage}{0.5\textwidth}
        \centering
        \includegraphics[width=1\textwidth]{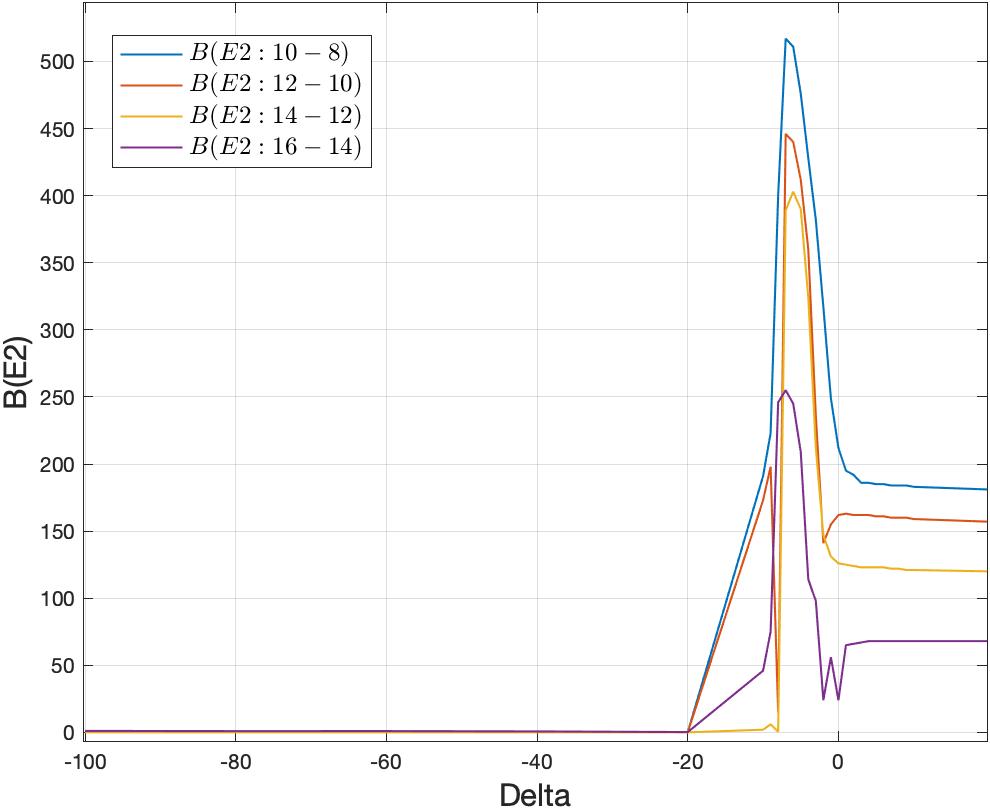}
        \caption{B(E2) 10, 12, 14, 16.}
        \label{f34}
        \end{minipage}
    \end{figure}

    \clearpage
 
    \subsection{SET 4}
    \subsection*{CASE - 2}
    Global view: Pair D ($p_{1/2}-p_{3/2}$) shifted by an amount $\Delta$ relative to pair D ($f_{5/2}-f_{7/2}$).
    \begin{figure}[H]
    \centering
        \begin{minipage}{0.5\textwidth}
        \centering
        \includegraphics[width=1\textwidth]{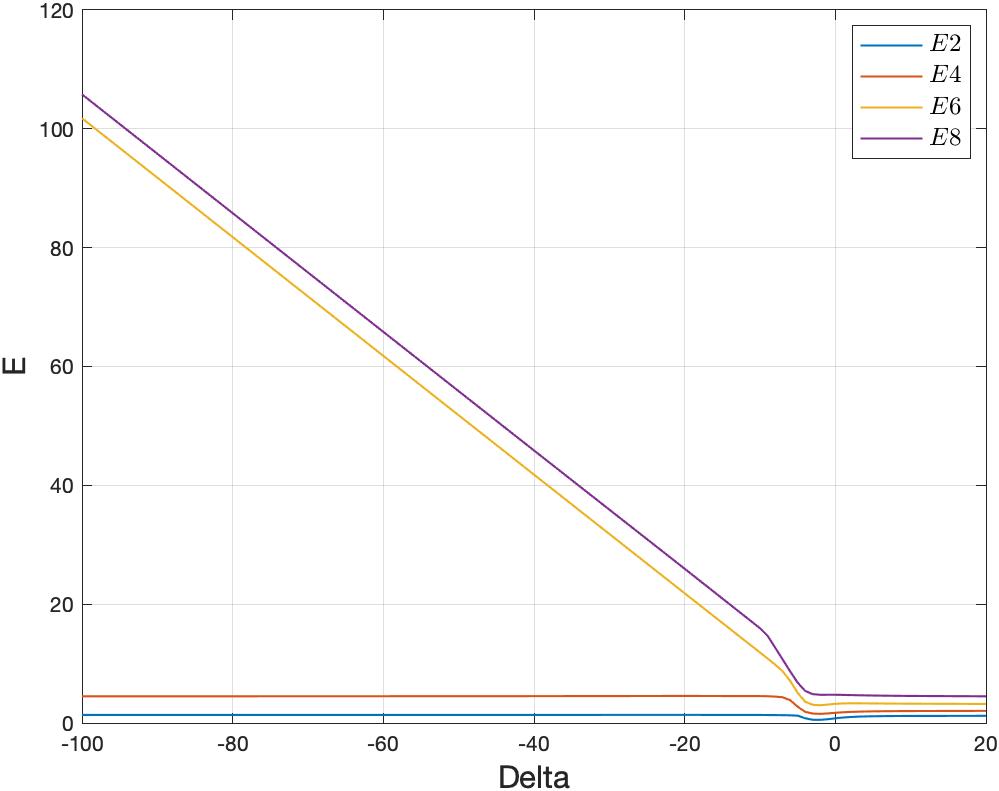}
        \caption{E2, E4, E6, E8.}
        \label{f41}
        \end{minipage}%
        \begin{minipage}{0.5\textwidth}
        \centering
        \includegraphics[width=1\textwidth]{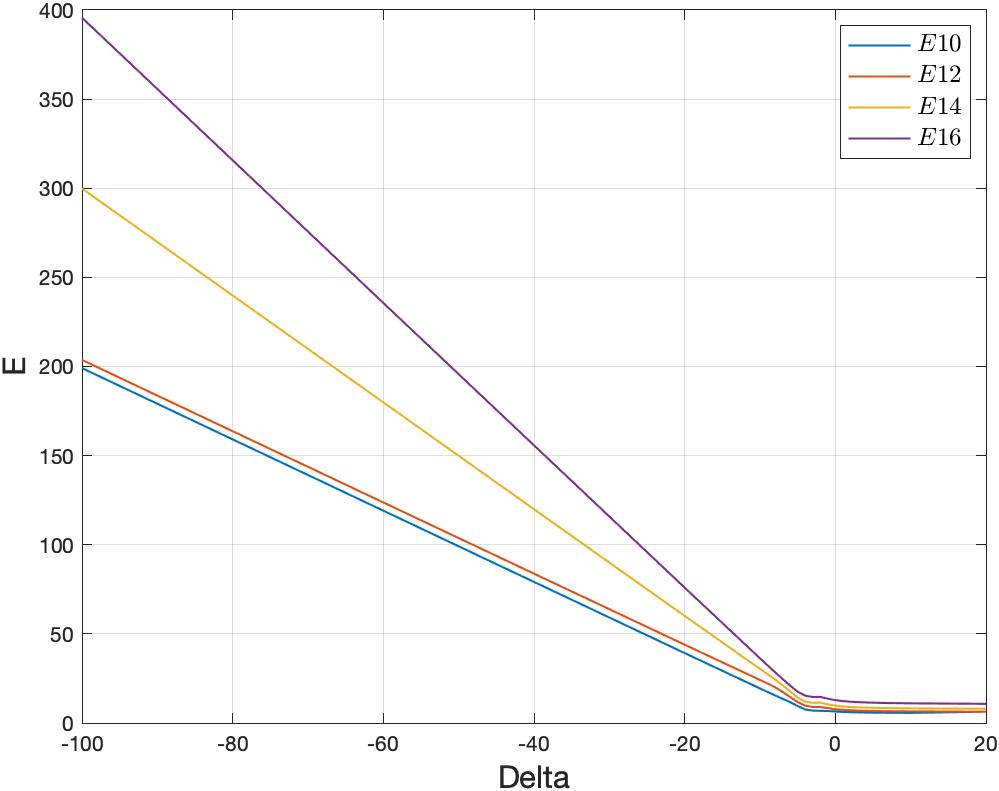}
        \caption{E10, E12, E14, E16.}
        \label{f42}
        \end{minipage}
    \end{figure}
    
    \begin{figure}[H]
    \centering
        \begin{minipage}{0.5\textwidth}
        \centering
        \includegraphics[width=1\textwidth]{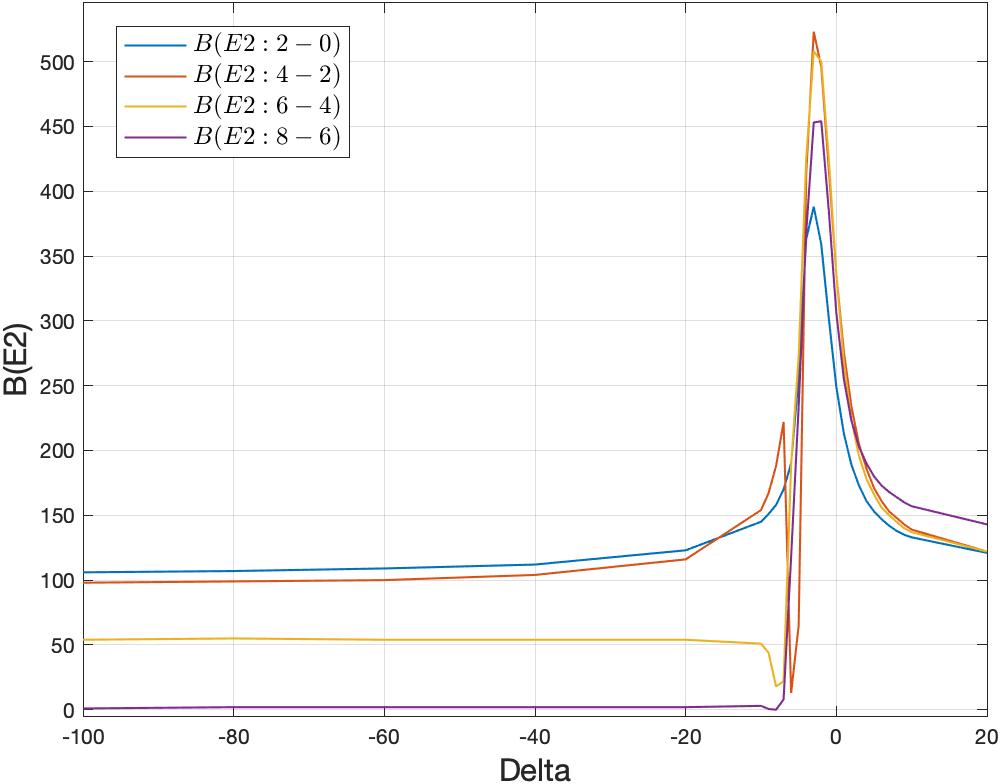}
        \caption{B(E2) 2, 4, 6, 8.}
        \label{f43}
        \end{minipage}%
        \begin{minipage}{0.5\textwidth}
        \centering
        \includegraphics[width=1\textwidth]{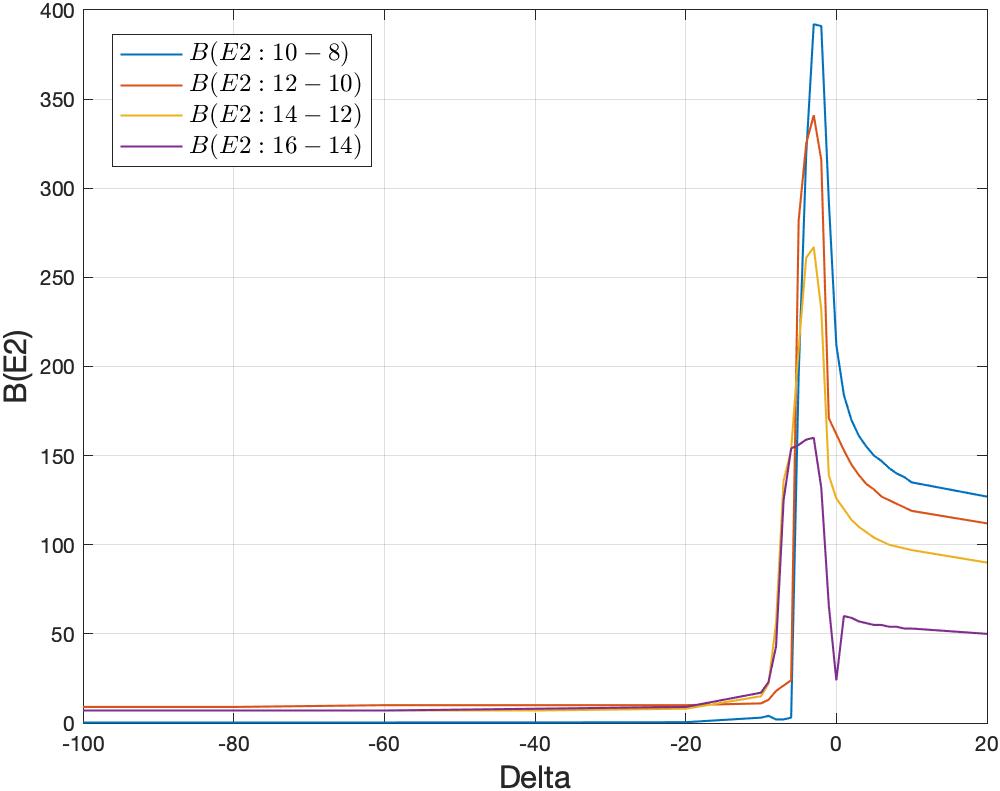}
        \caption{B(E2) 10, 12, 14, 16.}
        \label{f44}
        \end{minipage}
    \end{figure}

    \clearpage

\section{Overview of All the Results}
    A casual view of all the figures gives on the impassion that all the curves look very similar, be it energy levels or B(E2)'s, be it CASE - 1 or CASE - 2. There are gradual changes with increasing positive $\Delta$ in most cases. For $\Delta$ becoming more negative things change smoothly for a while but then there are sudden turn overs. This is undoubtedly due to the fact that there will be crossovers of single particle energies.

    For CASE - 1 we note with GXPF1A the 0$f_{5/2}$ is 4.4870 MeV above 0$f_{7/2}$ so with $\Delta$ lowers that -4.4870 we expect strange things to happen. In the first 4 figures we indeed see down turnovers at about that value of $\Delta$.
    
    For CASE - 2 we note that 0 1$p_{3/2}$ is 2.9447 MeV above 0$f_{7/2}$ so we we expect some sudden changes for $\Delta$ about -3 MeV. In contrast to CASE - 1 here we see sudden upturns as we further decrease $\Delta$. For the B(E2)'s there is a downturn for CASE - 2.

    We will give a more detailed discussion in the next subsection. We will focus on positive $\Delta$.

    \subsection{SET 1. CASE - 1}
    In Fig \ref{f11} where we shift pair B ($p_{1/2}-f_{5/2}$) relative to pair A we note that the excitation energies E2, E4, E6, and E8 slowly come down was we go from $\Delta$ = 0 to 20. However the overall spectrum seems not to change very much. 
    
    In Fig \ref{f12} we have E10, E12, E16 and E16 also coming down with increasing $\Delta$ but at a faster rate than for the lower J's in Fig 1. Again, the overall spectrum seems not to change very much.

    In Fig \ref{f13} The B(E2)s also come down with an increase in $\Delta$ but at a faster rate than the energies. This is probably due to the fact that when $\Delta$ increases there is less configuration mixing and hence less collectivity.
    
    In Fig \ref{f14} We show a striking difference in the B(E2) behavior
    for $J=10, 12, 14, 16$ when compared with the case in Fig \ref{f13} for $J=$2, 4, 6, and 8. Now the curves for positive $\Delta$ are much
    flatter-much less variation with increasing $\Delta$. A flat curve suggests configuration mixing , which is on the decease as
    $\Delta$ is increasing , is not so important. The higher the spin
    the more important are the single particle orbitals with high J
    e.g. $f_{7/2}$.
    
    There have been previous discussion of the band structure
    of $^{48}$Cr and the fact the the higher J states do not belong to
    the same "band" as the lower ones. The general consensus is
    that for the higher states there is an alignment of some of the
    nucleons along the rotation axis. This shows up more
    dramatically in the B(E2)s rather than the energies.

    \subsection{SET 2. CASE - 2}
    As shown in Fig \ref{f21} when we shift ($p_{1/2}-p_{3/2}$) to more
    positive values we find that the excitation energies go slightly
    up although the entire spectrum does not change very much. This is in contrast to the case of Section1 where the energies
    went slightly down. AN exception is $J=8^{+}$which is surprisingly flat.
    
    In Fig \ref{f22} we go to higher spin and on the whole the energies
    decrease with increasing $\Delta$. This is another indication that
    perhaps the high J states do not belong to the same band as the
    lower ones. For the first tile we see a crossover near $\Delta$ = 20
    with $J=12^{+}$starting to come down below $J= 10^{+}$.
    
    In Fig \ref{f23} we come back to B(E2)s. They decree with
    increasing $\Delta$ . This is not so different from The qualitative
    behavior in Fig \ref{f13}. This is again due to the fact that for
    increasing $\Delta$ there is less configuration mixing and hence
    less collectivity.
    
    In Fig \ref{f24}. We look at B(E2)s for $J=$10, 12, 14 and 16. There is a
    decrease with increasing $\Delta$ but not as severe as for $J=$2, 4, and 6. For the $J= 16^{+}$ to 14$^{+}$transition the curve is very flat.
    For such a high spin you need high $J$ single orbits to construct
    the state, so the reducing of contributions from $p_{3/2}$ and $p_{1/2}$
    is not so important.

    \subsection{Global view SET 3. CASE - 1}
    Here we will put more emphasis on negative $\Delta$. In Fig \ref{f31}
    we see large increase in the excitation energies as we say decrease $\Delta$ form +5 to -5. We could argue that at $\Delta$ =-5
    we have a lot of single particle orbitals close to together and
    this increases the pairing so that J=0 drops down a lot relative
    to $J =$ 2, 4, 6, and 8. For $\Delta$ even more negative the low $J$
    states are dominated by the $f_{5/2}$ and $p_{1/2}$ orbitals. As one
    makes $\Delta$ more negative configuration for ($f_{7/2},p_{3/2}$)
    becomes less important so the curve flattens out.
    
    In Fig \ref{f32} ($J = 10, 12, 14, 16$) the behavior for negative $\Delta$ is
    completely different than for low $J$. Except for $J=10^{+}$, the
    other energy levels got up in a linear fashion with increasing
    negative $\Delta$. This is not difficult to understand. On needs
    the high $J$ $f_{7/2}$ orbital to construct these states and moving say
    $f_{5/2}$ below $f_{7/2}$ is equivalent to putting $f_{7/2}$ above $f_{5/2}$.
    
    For the B(E2)s in Figs \ref{f33} for $J =$ 2, 4, 6 and 8 the flatness is due
    to the reduced configuration mixing due to the wide
    separation of the $f_{7/2}$ form $f_{5/2}$.

    \subsection{Global view SET 4. CASE - 2}
    In this case the lowest orbits have the lowest spins i.e. $p_{3/2}$
    and $p_{1/2}$. It is even harder to make low lying high spin states
    in this case. It is therefore not surprising to see in Fig \ref{f41} that
    even lower $J$ state energies i.e. $J=$6 and 8 have a linear rise
    with increasing negative  $\Delta$.
    
    In more detail with he $p_{3/2}$ and $p_{1/2}$ orbitals the maximum spin we can have for 4 protons is $J_p$ = 2. via the configuration $p_{3/2}$ $p_{3/2}$ $p_{3/2}$ $p_{1/2}$, which is equivalent to $p_{3/2}^{-1} p_{1/2}$. The same is true for 4 neutrons. So with $J_p$ = 2, $J_n$ = 2 we can only make states up to $J = 4$. This explains why only $J=0, 2, 4$ remain at low wineries as we increase negative $\Delta$.
    
    And of course in Fig \ref{f42} we see for the same reason that all
    high $J$ state excitation energies (10, 12, 14, and 16) rise linearly
    with increasing negative  $\Delta$. 
    
    The flattening of the B(E2)
    curves in Fig \ref{f43} and Fig \ref{f44} is due to the reduction in
    configuration mixing as the space between SET 3. and SET 4.
    single particle energies widens. The enhancement near  $\Delta$
    =-5 is due to the fact that all single particle orbitals are close
    to each other so there is an enhancement of collectivity due to
    the increase of configuration mixing.

\section{Closing Remarks - Scaling Behavior}
    In Table \ref{t6} we show the ratio E(J)$\Delta$/E(J) for $\Delta$=1, 10 and 20. Note that although there are some
    fluctuations the ratios are similar. If the ratios for a given $\Delta$ were all the same we would have
    perfect scaling. In that idealized situation we would get identical spectra for any finite $\Delta$ with that
    of $\Delta$=0 by multiplying the entire matrix for that $\Delta$ for that by a constant. Thus, is a
    phenomenological approach. If we limited ourselves to fitting the spectra of the 16 yrast states of
    $^{48}$Cr, we would have an infinite number of choices of combinations of 2 body matrix elements
    and single particle energies which would yield the same results. In truth as seen in Table \ref{t6} the
    ratios are not exactly the same but they are close enough to the idealized situation so that a large
    range of choices would lead to equally good results for these spectra. Of course if we expanded
    the data i.e. included other states. the result would be different. In Table \ref{t7} we compare the
    original spectrum of GXPF1A with that for $\Delta$=20 multiplied by a renormalization factor 1.2. This
    renormalization factor multiplies the entire matrix including the $\Delta$=20 single particle energies.
    We see that the spectra are reasonably close- it would be hard to prefer one to the other. However
    the single particle energies are vastly different. Originally the 0$f_{5/2}$ and 1$p_{1/2}$ are 7.241 and
    4.487 MeV above 0$f_{7/2}$. Now they are 27.241 and 24.487 MeV above the 0$f_{7/2}$ orbit.
    
    When one makes truncation in the PF shell by dropping orbits it is more natural to drop the spin
    orbit partners 0$f_{5/2}$ and 1$p_{1/2}$ than it is to drop the 2 p-shell orbits. This was done by Zamick et
    al. \cite{13} in the context of quadrupole moments and B(E2)'s. They studied the effects of dropping
    spin-orbit partners 0$f_{5/2}$ and 1 $p_{1/2}$ on these electromagnetic properties. In the present context
    this is equivalent to setting $\Delta$to infinity. To a large extent the results of the truncated calculations
    could be put into line with the full calculations by enlarging the effective charges in the former
    when the FPD6 interaction is used. The ratio full to truncated for Q(2$^+$) 2 , Q(4$^+$) 2 ,B(E2, 2$^+$ $\rightarrow$
    0$^+$) and B(E2 , 4$^+$ $\rightarrow$ 2$^+$) were all very close to 1.4.
    The possibility of scaling behavior is intriguing and will be further investigated in the near
    future.

\clearpage

\section*{ACKNOWLEDGEMENTS}

P C Srivastava acknowledges a  research grant from SERB (India), 
CRG/2019/000556 and Kalam cluster at Physics Department, IIT-Roorkee. 
C. Fan acknowledges supports from Aresty Research Center.


\vspace{0.8cm}


\begin{thebibliography}{9}
\bibitem{1} M. Honma, T. Otsuka, B. A. Brown, and T. Mizusaki,
Phys. Rv. C 65, 061301 (R), (2002); Phys. Rev. C 69, 034335 (2004).

\bibitem{2} W.A. Richter, M.G. Van Der Merwe, R.E. Julius and
B.A. Brown, Nucl. Phys. A 523, 325 (1991).

\bibitem{3} E. Caurier, G. Martinez-Pinedo, F. Nowacki, A. Poves,
and A. P. Zuker, Rev. Mod. Phys. 77, 427 (2005).

\bibitem{4} E. Caurier, A.P. Zuker, A. Poves and C. Martinez-Pinero,
Phys. Rev. C50, 225 (1994).

\bibitem{5} E. Caurier, F. Nowacki , A. P. Zuker , G. Martinez-Pinedo, A. Poves, and J. Retamosa , Nuclear Physics A 654, 747 (1999).

\bibitem{6} V. Kumar, P. C. Srivastava and A. Kumar,
Acta Physica Polonica B 51, 961 (2020). 


\bibitem{7} Kenji Hara, Yang Sun, and Takahiro Mizusaki,
Phys. Rev. Lett. 83, 1922 (1999).


\bibitem{8} F. Brandolini and C. A. Ur,
Phys. Rev. C 71, 054316 (2005).


\bibitem{9} E. Caurier, J. L. Egido, G. Martinez-Pinedo, A. Poves, J. Retamosa, L. M. Robledo, and A. P. Zuker
Phys. Rev. Lett. 75, 2466 (1995).


\bibitem{10} Zao-Chun Gao, Mihai Horoi, Y. S. Chen, Y. J. Chen, and Tuya,
Phys. Rev. C 83, 057303 (2011).


\bibitem{11} R. A. Herrera and C. W. Johnson,
Phys. Rev. C 95, 024303 (2017). 

\bibitem{12} Y.Y. Sharon, N. Benczer-Koller, G. J. Kumbartzki,
L. Zamick, R. F. Casten, Nuclear Physics A 980, 131 (2018). 

\bibitem{13} L. Zamick, Y.Y.Sharon, S.J.Q. Robinson and M. Harper,
Phys. Rev. C 91, 064321 (2015).


\end{thebibliography}
\end{document}